# Functional methods in the theory of magnetoimpurity states of electrons in quantum wires

A.M. Ermolaev,  G.I. Rashba

Department of Theoretical Physics, V.N. Karazin Kharkiv National University,
4 Svobody sq, 61077, Kharkiv, Ukraine
Email: `georgiy.i.rashba@univer.kharkov.ua`

Functional methods are used to study magnetoimpurity states of electrons in nanostructures. The Keldysh formalism is applied to these states. The theory is illustrated using a quantum wire sample with impurity atoms capable of localizing electrons in a magnetic field. The characteristics of magnetoimpurity states of electrons in the wire are calculated using the model of a Gaussian separable potential.

## 1. Introduction

Impurity states of electrons (local and resonant) in alloyed metals and semiconductors have been known for a long time (see, for instance, [1-4]). It is well known, in particular, that shallow and narrow impurity potential well in a three-dimensional conductor cannot localize an electron. In one-dimensional and two-dimensional cases localization is possible with any well intensity [5].

In the presence of a quantizing magnetic field the situation changes. The motion of an electron in a massive conductor in a magnetic field is similar to a one-dimensional case, and in the one-dimensional case the electron is localized in a potential well of any degree of intensity [5]. The spatial inhomogeneity of the alloyed conductor results in removal of degeneration of the Landau levels over the position of "Landau oscillator". Impurity levels split off from every Landau level. Levels split off from the lower Landau level are local and levels split off from the higher levels fall within an area of continuous spectrum and are resonant [6,7]. Such levels caused by mutual influence of donor-impurity atoms and magnetic field on electrons are called magnetoimpurity levels, and the respective states of electrons are called magnetoimpurity states.

The magnetoimpurity states of electrons create the beats in the de Haas-van Alfven [6,7] and in the Shubnikov-de Haas effects and they cause the linear growth of magnetic resistance of metals with closed Fermi surfaces along with the growth of magnetic field [8,9]. These states result in appearance of new branches of collective modes in massive and low-dimensional conductors: electromagnetic [10-12], sound [13], magnetoplasma [14], and electronic spin waves [15].

The elaborated methods of studying impurity states are based on the Schrödinger's equation and on the quantum Green's functions. In the meantime, functional methods derived in quantum field theory are widely adopted in the solid-state theory [16,17]. In conjunction with the Keldysh technique [18,19] they become a powerful tool of research in condensed matter physics [17,20,21]. Here we will specify how the methods of functional differentiation and integration using Grassmann variables can be used to calculate the characteristics of magnetoimpurity states of electrons in massive conductors and nanostructures. The formalism set out in section 2 and a model presented in section 3 are used in section 4 to calculate the characteristics of magnetoimpurity states of electrons in quantum wires.

## 2. Formalism

The functional approach to the Keldysh formalism has recently been the subject of a comprehensive monograph [17] including applications in disordered conductors. As far as the authors are aware, the potential of functional methods in the theory of magnetoimpurity states of electrons in conductors [6,11] has not been studied yet. In this section, we will describe how the formalism of works [17,22] is applied to arbitrary magnetic field and magnetoimpurity states of electrons on isolated impurity atoms. This theory is applicable to solid conductors and two-dimensional electron gas. The general theory of magnetoimpurity states of electrons in two-dimensional electron systems has been developed in the series of works by Azbel, Gredeskul, Avishai and others (see review [23]) using a different method of calculation. The advantages of functional methods compared with standard methods consist in the fact that transformation from operators to classical objects allows us to simplify, to speed up and to make calculations more visual. The universal language of the functional approach permits to describe a lot of phenomena in equilibrium and non-equilibrium systems using the unique method. In particular, Keldysh formalism permits us to study the influence of magnetoimpurity states of electrons upon non-equilibrium systems properties.

Let us write the Hamiltonian of electrons as $\hat{H} = \hat{H}_0 + \hat{V}$, where $\hat{H}_0$ accounts for the magnetic field, and $\hat{V}$ – the interaction with a random external field. Similarly to works [17,22], we con-sider that it is switched "on" at the initial time $t = -\infty$, when the system of electrons was yet in the state of equilibrium. Then the matrix function of Green electrons [17,22] can be written as

$$iG_{12} = \left\langle T_{C_t}\left(\hat{\psi}_{1H}\hat{\psi}_{2H}^+\right)\right\rangle, \qquad (1)$$

where indices 1 and 2 correspond to variables $\mathbf{r},\alpha,t$ ($\mathbf{r}-$ is a position vector, $\alpha = \pm 1$ – spin variable, $t$ – time), $\hat{\psi}_{1H}$ and $\hat{\psi}_{2H}^+$ – Heisenberg operators of destruction and creation of electrons, $T_{C_t}$ – the symbol of chronological ordering of operators along the Keldysh-Schwinger time contour of $C_t = C_+ + C_-$ (figure 1,a), the angle brackets indicate Gibbs averaging with the Hamiltonian $\hat{H}_0$. The reduced Planck constant is set to unity. In Eq. (1) matrix indices $\pm$ in two-dimensional Keldysh space are omitted. In the component form the Green function (1) takes the form

$$G_{12} = \begin{pmatrix} G_{12}^{++} & G_{12}^{+-} \\ G_{12}^{-+} & G_{12}^{--} \end{pmatrix}. \qquad (2)$$

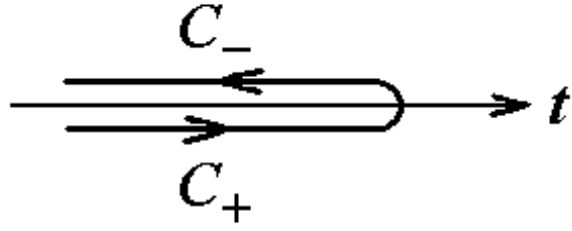

a)

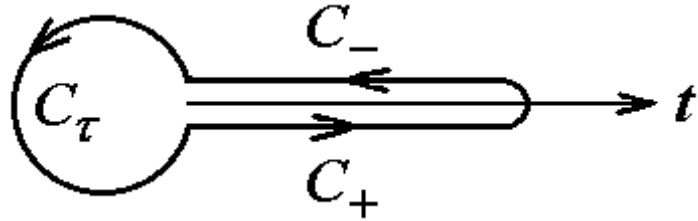

b)

Figure 1. Keldysh-Schwinger time contour (a) and extended contour (b).

Here $G_{12}^{+-}$, for instance, indicates that $t_1 \in C_+$ and $t_2 \in C_-$. It is convenient to rewrite the operators $\hat{\psi}_{1S}$ and $\hat{\psi}_{2S}^+$ in Eq.(1) in the Schrödinger representation. Then,

$$iG_{12} = \left\langle T_{C_t}\left(\hat{\psi}_{1S}\hat{\psi}_{2S}^+\hat{U}\right)\right\rangle, \qquad (3)$$

where

$$\hat{U}(t,t_0) = T_{C_t} \exp\left[-i\int_{t_0}^{t} dt' \hat{H}(t')\right]$$

is the evolution operator along the contour $C_t$.

By using the standard procedure [16,17], let us present the Green function (3) as a path integral along Grassmann fields $\psi_1$, $\psi_2^*$, associated with the field operators. For that we introduce a Gibbs exponent $\exp(-\beta\hat{H}_0)$ ($\beta$ – reverse temperature) under the chronological product operator. This is achieved by using extended contour $C = C_t + C_\tau$ (figure 1,b) and the time-ordering operator $T_C$ along this contour. It is assumed that on the part of the contour $C_\tau$ the Hamiltonian of electrons is equal to $\hat{H}_0$. Thus the Green function (3) looks as the following

$$iG_{12} = \int D\psi^* \int D\psi\, \psi_1 \psi_2^* \exp\left\{i\oint_C dt\left[\sum_\alpha \int d\mathbf{r}\, \psi_\alpha^*(\mathbf{r},t) i\frac{\partial}{\partial t}\psi_\alpha(\mathbf{r},t) - H(\psi^*,\psi)\right]\right\}. \quad (4)$$

Function $H(\psi^*,\psi)$ is derived from the operator $H(\hat{\psi}^+,\hat{\psi})$ by replacing the field operators with the Grassmann variables.

The Green functions can be derived by functional differentiation of the generating functional

$$Z[J,J^*] = \int D\psi^* \int D\psi\, \exp\{i\oint_C dt\left[\sum_\alpha \int d\mathbf{r}\, \psi_\alpha^*(\mathbf{r},t) i\frac{\partial}{\partial t}\psi_\alpha(\mathbf{r},t) - H(\psi^*,\psi) + \right.$$

$$\left. + \sum_\alpha \int d\mathbf{r}\, \psi_\alpha^*(\mathbf{r},t) J_\alpha(\mathbf{r},t) + \sum_\alpha \int d\mathbf{r}\, J_\alpha^*(\mathbf{r},t) \psi_\alpha(\mathbf{r},t)\right]\} \quad (5)$$

by sources $J$, $J^*$ of Fermi field. In particular, function (4) is equal to

$$iG_{12} = \left(\frac{\delta^L}{i\delta J_1^*} \frac{\delta^R}{i\delta J_2} W[J,J^*]\right)_{\substack{J=0 \\ J^*=0}}, \quad (6)$$

where $W = \ln Z$ – is the generating functional for connected Green functions, indices $L$ and $R$ mark the left and right functional derivatives over the Grassmann variables $J^*$, $J$.

Let us transform Eq.(5) into $(\kappa,\alpha,t)$–representation ($\kappa$ – is a set of orbital quantum electron numbers in the magnetic field) and calculate the free generating functional in the magnetic field and field of sources:

$$Z_0[J,J^*] = \int Da^* \int Da \exp\left\{i\oint_C dt \sum_{\kappa\alpha}\left[a^*_{\kappa\alpha}(t)\left(i\frac{\partial}{\partial t}-\varepsilon_{\kappa\alpha}\right)a_{\kappa\alpha}(t)+\right.\right.$$
$$\left.\left.+J^*_{\kappa\alpha}(t)a_{\kappa\alpha}(t)+a^*_{\kappa\alpha}(t)J_{\kappa\alpha}(t)\right]\right\}. \quad (7)$$

Here $a_{\kappa\alpha}$, $a^*_{\kappa\alpha}$, $J_{\kappa\alpha}$, $J^*_{\kappa\alpha}$ – are expansion coefficients of Grassmann variables $\psi$, $\psi^*$, $J$, $J^*$ over the Landau basis, $\varepsilon_{\kappa\alpha}$ – energy of electron in the magnetic field, calculated from the chemical potential. The functions $a(t)$ and $a^*(t)$ meet the boundary conditions:

$$\begin{aligned}
a_+(\infty) &= a_-(\infty), & a^*_+(\infty) &= a^*_-(\infty), \\
a_-(-\infty) &= a_\tau(0), & a^*_-(-\infty) &= a^*_\tau(0), \\
a_+(-\infty) &= -a_\tau(\beta), & a^*_+(-\infty) &= -a^*_\tau(\beta).
\end{aligned} \quad (8)$$

Indices $+$, $-$, $\tau$ are associated with the parts $C_+$, $C_-$, $C_\tau$ of the expanded contour $C$.

In Eq.(7) let us split the coefficients $a = a_0 + a'$, $a^* = a_0^* + a'^*$, where $a_0$ and $a_0^*$ are such that the linear terms over Grassmann variables $a'$, $a'^*$ in the exponent index drop out. Therefore it is necessary that the functions $a_0^+$, $a_0^-$, $a_0^\tau$ satisfy the system of equations

$$\left(i\frac{\partial}{\partial t}-\varepsilon\right)a_0^+ = -J^+, \quad \left(i\frac{\partial}{\partial t}-\varepsilon\right)a_0^- = J^-, \quad \left(\frac{\partial}{\partial \tau}+\varepsilon\right)a_0^\tau = 0, \quad (9)$$

and conditions (8). We placed $J = J^+$ on the part of the contour $C_+$ and $J = -J^-$ on $C_-$. Variables $t$ and $\tau$ are related as $\tau = it$.

The solution of the system of equations (9) can be written as

$$a^0_{\kappa\alpha}(t) = -\int_{-\infty}^{\infty} dt'\, \overset{0}{G}_{\kappa\alpha}(t-t')J_{\kappa\alpha}(t'),$$

where $\overset{0}{G}$ – is a matrix Green function of electrons in the magnetic field. Its components are equal to

$$\overset{0}{G}{}^{++}_{\kappa\alpha}(t) = i\left[\Theta(t)(f_{\kappa\alpha}-1)+\Theta(-t)f_{\kappa\alpha}\right]\exp(-i\varepsilon_{\kappa\alpha}t),$$

$$\overset{0}{G}{}^{+-}_{\kappa\alpha}(t) = i f_{\kappa\alpha}\exp(-i\varepsilon_{\kappa\alpha}t),$$

$$\overset{0}{G}{}^{-+}_{\kappa\alpha}(t) = i(f_{\kappa\alpha}-1)\exp(-i\varepsilon_{\kappa\alpha}t),$$

$$\overset{0\;--}{G_{\kappa\alpha}}(t) = i\left[\Theta(t)f_{\kappa\alpha} + \Theta(-t)(f_{\kappa\alpha} - 1)\right]\exp(-i\varepsilon_{\kappa\alpha}t),$$

where $\Theta$ and $f$ – are the Hevyside and the Fermi functions respectively.

The path integral (7) over the Grassman variables $a'^*, a'$ is the Gaussian integral. It can be calculated according to the rule [16,17]

$$\int Da'^* \int Da' \exp(-a'^* Aa') = \det A,$$

here $A$ – is a non-singular matrix being irrelevant for our derivation. An important factor in the generating functional (7) is equal to

$$Z_0[J, J^*] = \exp\left(-i\sum_{12} J_1^* \overset{0}{G}_{12} J_2\right). \qquad (10)$$

Here $1 = (\kappa_1, \alpha_1, t_1)$, $\sum_1 = \sum_{\kappa_1\alpha_1}\int_{-\infty}^{\infty}dt_1$. The summation symbol over the Keldysh indices $\pm$ in Eq. (10) is omitted.

The following Hamiltonian $\hat{V}$ accounts for the interaction of electrons with impurity atoms randomly distributed in a sample:

$$\hat{V} = \sum_{\kappa_1\kappa_2\alpha}\langle\kappa_1|u|\kappa_2\rangle\hat{a}^+_{\kappa_1\alpha}\hat{a}_{\kappa_2\alpha},$$

where $\hat{a}_{\kappa\alpha}$ and $\hat{a}^+_{\kappa\alpha}$ – are destruction and creation operators of electrons in state $(\kappa, \alpha)$, $u$ – the energy of electron in the field of impurity atoms. Since $V = 0$ on a part of the contour $C_\tau$, the integral in exponent index (5) is equal to

$$\oint_C dt\, V(a^*, a) = \sum_{12} U_{12} a_1^* \sigma_3 a_2, \qquad (11)$$

where $U_{12} = \langle\kappa_1|u|\kappa_2\rangle\delta_{\alpha_1\alpha_2}\delta(t_1 - t_2)$, $\sigma_3$ – the third Pauli matrix.

From Eq.(5) it is clear that the full generating functional can be derived from the free functional (10) by operator action

$$\exp\left(-i\int_{-\infty}^{\infty}dt\, V\right),$$

obtained from Eq.(11) as a result of replacement

$$a^*_{\kappa\alpha}(t) \to \frac{\delta^R}{i\delta J_{\kappa\alpha}(t)}, \qquad a_{\kappa\alpha}(t) \to \frac{\delta^L}{i\delta J^*_{\kappa\alpha}(t)}.$$

Consequently, the generating functional (5) is equal to

$$Z[J,J^*] = \exp\left(-i\sum_{12} U_{12} \frac{\delta^R}{i\delta J_1} \sigma_3 \frac{\delta^L}{i\delta J_2^*}\right) \exp\left(-i\sum_{34} J_3^* \overset{0}{G}_{34} J_4\right). \tag{12}$$

In exponent indices it is necessary to execute summing over the Keldysh indices. Thus, the procedure of calculation of the Green function of the system can be done by calculation of functional derivative in equations (6) and (12). This procedure is easier than the method of local perturbations using operator formalism.

Equation (12) enables us to write the generating functional in the form of a series over perturbation $V$. Green function (6) is also presented in the form of a series which coincides with the series derived by using the diagram technique and the Wick's theorem [16]. The first order correction to the generating functional $W$ is equal to

$$W_1[J,J^*] = \sum_{12} U_{12}\left(\overset{0\,++}{G}_{21} - \overset{0\,--}{G}_{21}\right) - i\sum_{12} U_{12}\sum_{34}\left[J_3^{*+} J_4^+\left(\overset{0\,+-}{G}_{31}\overset{0\,-+}{G}_{24} - \overset{0\,++}{G}_{31}\overset{0\,++}{G}_{24}\right) + \right.$$

$$+ J_3^{*+} J_4^-\left(\overset{0\,+-}{G}_{31}\overset{0\,--}{G}_{24} - \overset{0\,++}{G}_{31}\overset{0\,+-}{G}_{24}\right) + J_3^{*-} J_4^+\left(\overset{0\,--}{G}_{31}\overset{0\,-+}{G}_{24} - \overset{0\,-+}{G}_{31}\overset{0\,++}{G}_{24}\right) +$$

$$\left. + J_3^{*-} J_4^-\left(\overset{0\,--}{G}_{31}\overset{0\,--}{G}_{24} - \overset{0\,-+}{G}_{31}\overset{0\,+-}{G}_{24}\right)\right].$$

Using equation (6) we obtain the first order corrections to the components of matrix Green function (2):

$$\overset{1\,++}{G}_{12} = \sum_{34} U_{34}\left(\overset{0\,+-}{G}_{13}\overset{0\,-+}{G}_{42} - \overset{0\,++}{G}_{13}\overset{0\,++}{G}_{42}\right),$$

$$\overset{1\,+-}{G}_{12} = \sum_{34} U_{34}\left(\overset{0\,+-}{G}_{13}\overset{0\,--}{G}_{42} - \overset{0\,++}{G}_{13}\overset{0\,+-}{G}_{42}\right),$$

$$\overset{1\,-+}{G}_{12} = \sum_{34} U_{34}\left(\overset{0\,--}{G}_{13}\overset{0\,-+}{G}_{42} - \overset{0\,-+}{G}_{13}\overset{0\,++}{G}_{42}\right),$$

$$\overset{1\,--}{G}_{12} = \sum_{34} U_{34}\left(\overset{0\,--}{G}_{13}\overset{0\,--}{G}_{42} - \overset{0\,-+}{G}_{13}\overset{0\,+-}{G}_{42}\right).$$

The diagrams for these corrections are presented in [24]. They differ from the usual diagrams of the cross technique [25] by additional indices $\pm$ at the ends of lines.

Here we restrict ourselves to a selective summing of diagrams with one cross for retarded Green's function $G = G^{++} - G^{+-}$ of electrons in a nanostructure averaged over impurity configurations. Such approximation allows us to consider accurately the amplitude of electron scattering by isolated impurity atoms with a small concentration of such atoms. Let's choose the scattering potential in the form

$$\hat{V} = \sum_j |\eta_j\rangle u_0 \langle \eta_j|, \qquad (13)$$

where $|\eta_j\rangle\langle\eta_j|$ – is the projection operator on vector $|\eta_j\rangle$, $u_0$ – a constant, index $j$ numbers impurity atoms. We deem that function $\eta(\mathbf{r}) = \langle \mathbf{r}|\eta\rangle$ is equal to

$$\eta(r) = \left(\sqrt{\pi}a\right)^{-1} \exp\left(-\frac{r^2}{2a^2}\right),$$

where $a$ – is a constant. Transformation of Eq.(13) to the sum of delta functions $\upsilon_0 \delta(\mathbf{r} - \mathbf{r}_j)$ is performed by replacement

$$4\pi \lim_{\substack{a \to 0 \\ u_0 \to \infty}} \left(a^2 u_0\right) = \upsilon_0.$$

In case of potential (13) the sum of diagrams with one cross for mean Green function in $(\kappa, \alpha, \varepsilon)$-presentation is equal to $G = G_0 + G_0 T G_0$, where value

$$T_\alpha(\varepsilon) = u_0 n_i \left(1 - u_0 \sum_\kappa \frac{|\langle \kappa|\eta\rangle|^2}{\varepsilon - \varepsilon_{\kappa\alpha}}\right)^{-1} \qquad (14)$$

is proportional to the scattering amplitude of electrons, $n_i$ – density of impurity atoms.

### 3. Quantum wire model

The model of quantum wire convenient for calculations is a system of electrons in a two-dimensional conductor on which the confinement potential is placed restricting their motion in one direction. A sample of quantum wire is a two-dimensional electron gas at the border of Si and $SiO_2$ in the system like metal-dielectric-semiconductor or in heterostructure $GaAs/Al_xGa_{1-x}As$ with confinement potential created using electron beam lithography. In such systems electrons move freely in one direction and their motion in other directions is restricted.

Let us select vector potential of magnetic field H perpendicular to plane $(x, y)$ occupied by a two-dimensional electron gas in the form of $\vec{A} = (0, Hx, 0)$. Confinement potential is assumed to be parabolic $m\omega_0^2 x^2/2$. Here $m$ – is the effective mass of electron, $\omega_0$ – is the potential parameter. Then the orbital wave function of electron's stationary state looks as usual [5]

$$\psi_{nk_y}(x,y) = \left(\sqrt{\pi}\, 2^n n!\, lL\right)^{-\frac{1}{2}} \exp\left[-\frac{1}{2}\left(\frac{x-x_0}{l}\right)^2\right] H_n\left(\frac{x-x_0}{l}\right) \exp(ik_y y), \qquad (15)$$

but its parameters are renormalized. Here $n$ and $k_y$ – are the oscillator quantum number and the projection of electron momentum on wire axis $y$, $l = \left(1/m\omega\right)^{1/2}$, $\omega = \left(\omega_0^2 + \omega_c^2\right)^{1/2}$ – hybrid frequency ($\omega_c$ – electron's cyclotron frequency), $x_0 = -\omega_c k_y / (m\omega^2)$, $L$ – length of wire, $H_n$ – Hermite polynomial. Electron's energy in state (15) is equal to

$$\varepsilon_{nk_y\sigma} = \omega\left(n+\frac{1}{2}\right) + \frac{k_y^2}{2M} + \sigma\mu_B H, \qquad (16)$$

where $M = m\left(\omega/\omega_0\right)^2$, $\mu_B$ – electron spin magnetic moment, $\sigma = \pm 1$ – spin quantum number. Density of electron states with spectrum (16) is equal to

$$\nu_\sigma(\varepsilon) = \sqrt{\frac{M}{2}}\, \frac{L}{\pi} \sum_n \frac{\Theta(\varepsilon - \varepsilon_{n\sigma})}{\sqrt{\varepsilon - \varepsilon_{n\sigma}}},$$

where $\varepsilon_{n\sigma} = \omega\left(n+\frac{1}{2}\right) + \sigma\mu_B H$ – are Landau levels of electron in the wire.

### 4. Magnetoimpurity states of electrons in quantum wires

In quantum wires even separate impurity atoms significantly affect the properties of conduction electrons. In this section we will derive the equation of I. Lifshitz [26] for the spectrum of impurity states of electrons in quantum wire located in magnetic field. We approximate the potential of an impurity atom located at the center of coordinate system using convenient for calculations separable Gaussian potential (13).

Electron scattering operator with an impurity center looks like

$$T = \frac{|\eta\rangle u_0 \langle\eta|}{1 - u_0 D}, \qquad (17)$$

where

$$D_\sigma(\varepsilon + i0) = \sum_{nk_y} \frac{|\langle\eta|nk_y\rangle|^2}{\varepsilon - \varepsilon_{nk_y\sigma} + i0} = F_\sigma(\varepsilon) - i\pi g_\sigma(\varepsilon).$$

The operator poles in Eq.(17) correspond to impurity states of electrons. Impurity levels satisfy the equation of I. Lifshitz

$$1 - u_0 D_\sigma(\varepsilon + i0) = 0. \tag{18}$$

This equation can be obtained from the formula (3.63) in Ref. [27], if one would consider the self-energy function $\Sigma$, entering into this formula, in the linear approximation over $n_i$ and would take into account multiple scattering of electrons by isolated short-range impurity atoms in quantizing magnetic field. In this case the formula takes the following form in our notations

$$G_\sigma(\kappa, \varepsilon) = \left[\varepsilon - \varepsilon_{\kappa\sigma} - \frac{n_i u_0}{1 - u_0 D_\sigma(\varepsilon + i0)}\right]^{-1}.$$

In the linear approximation over $n_i$ this equation implies the relation between functions $G$, $G_0$ (presented in section 2) and multiple impurity scattering operator $T$. As it is noticed in Ref. [27], this is equivalent to replacement of the Born scattering amplitude of electrons by isolated impurity atom in Eqs. (3.61) and (3.67), which describe collision frequency of electrons, by $T$-matrix (see section 3.7 in [27]). The $T$-matrix accounts for the effects of potential and resonant electron scattering by impurity atoms. Replacing the resonant dominator $(1 - u_0 F_\sigma)^2 + (\pi u_0 g_\sigma)^2$ in function $\Sigma$ by the unity we obtain Born potential-scattering contribution into the electrons collision frequency. Near the poles of $T$-matrix satisfying Eq. (18) the amplitude is characterized by the Breit-Wigner resonances [5], which influence considerably on observed values [6-13,15].

Considering (13) and (15), we derive

$$\langle \eta | n k_y \rangle = \frac{2\sqrt{\pi} l}{\sqrt{\sqrt{\pi} 2^n n! (\alpha + 1) L}} \left(\frac{|\alpha - 1|}{\alpha + 1}\right)^{n/2} \exp\left[-\frac{1}{2} a^2 k_y^2 \left(1 + \frac{\alpha^2 \xi^2}{\alpha + 1}\right)\right] H_n\left(a k_y \xi \frac{\alpha^{3/2}}{\sqrt{|\alpha^2 - 1|}}\right),$$

where $\alpha = (l/a)^2$, $\xi = \omega_c/\omega$. As a result equation (18) takes the form

$$1 - u_0 \frac{2l}{\alpha + 1} \sum_n \int_{-\infty}^{\infty} dk_y \frac{\gamma^n}{\varepsilon - \varepsilon_{nk_y\sigma} + i0} \frac{1}{\sqrt{\pi} 2^n n!} \exp\left[-a^2 k_y^2 \left(1 + \frac{\alpha^2 \xi^2}{\alpha + 1}\right)\right] H_n^2\left(a k_y \xi \frac{\alpha^{3/2}}{\sqrt{|\alpha^2 - 1|}}\right) = 0, \tag{19}$$

where $\gamma = |\alpha - 1|/(\alpha + 1)$. Considering the identity

$$\frac{1}{\varepsilon - \varepsilon_{nk_y\sigma} + i0} = -i \int_0^\infty du \exp\left[iu\left(\varepsilon - \varepsilon_{nk_y\sigma} + i0\right)\right]$$

and calculating the integral included in (19), equation (19) can be written as

$$1 + i\frac{2^{3/2} u_0 l M^{1/2}}{\omega^{1/2}(\alpha+1)} \int_0^\infty dx \exp\left[ix\left(\frac{\varepsilon}{\omega} - \frac{1}{2} - \sigma\frac{\mu_B H}{\omega} + i0\right)\right]\left(1 - \gamma^2 e^{-2ix}\right)^{-1/2} \times$$

$$\times \left[ix + 2Ma^2\omega\left(1 + \frac{\alpha^2\xi^2}{\alpha+1}\right) - 4Ma^2\omega\frac{\alpha^3\xi^2}{(\alpha+1)^2}\frac{1}{e^{ix}+\gamma}\right]^{-1/2} = 0. \tag{20}$$

Separating the real and imaginary parts in this equation, we derive functions $F_\sigma(\varepsilon)$ and $g_\sigma(\varepsilon)$. The latter is equal to

$$g_\sigma(\varepsilon) = \frac{2^{3/2} l M^{1/2}}{\alpha+1} \sum_n \frac{\Theta(\varepsilon - \varepsilon_{n\sigma})}{\sqrt{\varepsilon - \varepsilon_{n\sigma}}} \frac{\gamma^n}{\sqrt{\pi} 2^n n!} \exp\left[-2Ma^2(\varepsilon - \varepsilon_{n\sigma})\left(1 + \frac{\alpha^2\xi^2}{\alpha+1}\right)\right] \times$$

$$\times H_n\left(\sqrt{2M}a\sqrt{\varepsilon - \varepsilon_{n\sigma}}\,\xi\frac{\alpha^{3/2}}{\sqrt{|\alpha^2-1|}}\right).$$

From equation (20) we understand that if $u_0 < 0$ in the spectrum of electrons there is a system of impurity levels $\varepsilon_{n\sigma}^{(r)}$ split off from Landau levels. A level below the border of continuous spectrum $\varepsilon_{0(-1)}$ is local, other levels are quasilocal. Their widths are equal to $\Gamma = \pi g / |F'|$, where the prime marks the derivative over energy taken in point $\varepsilon_{n\sigma}^{(r)}$. These levels result from a mutual impact of impurity atom and magnetic field on electrons. Therefore they are called magnetoimpurity states.

Equations (19) and (20) state that if $u_0 < 0$ then the position of local level $\varepsilon_{l\sigma}$ in zone $\varepsilon \leq \varepsilon_{0\sigma}$ is the root of the equation

$$1 - \frac{4\sqrt{\pi} M l |u_0|}{(\alpha+1)k_{0\sigma}} \exp\left(a^2 k_{0\sigma}^2 \rho^2\right)\left[1 - \Phi(ak_{0\sigma}\rho)\right] = 0, \tag{21}$$

where

$$k_{0\sigma} = \sqrt{2M(\varepsilon_{0\sigma} - \varepsilon)}, \qquad \rho = \sqrt{1 + \frac{\alpha^2\xi^2}{\alpha+1}},$$

$\Phi$ – is the probability integral [28]. If $ak_{0\sigma}\rho \ll 1$, from this equation we find the distance $\Delta_0$ between the local level and the border of the continuous spectrum:

$$\Delta_0 = \varepsilon_{0\sigma} - \varepsilon_{l\sigma} = \frac{8\pi Ma^2 u_0^2 \alpha}{(\alpha+1)^2} = \begin{cases} \dfrac{4\pi u_0^2 \alpha}{\varepsilon_0}, & \alpha \ll 1, \\ \dfrac{4\pi u_0^2}{\alpha\varepsilon_0}, & \alpha \gg 1. \end{cases} \tag{22}$$

Here $\varepsilon_0 = (2Ma^2)^{-1}$. In the extreme case $ak_{0\sigma}\rho \ll 1$ from equation (21) we derive

$$\Delta_0 = \frac{2|u_0|\sqrt{\alpha}}{(\alpha+1)\rho} = \begin{cases} 2|u_0|\sqrt{\alpha}, & \alpha \ll 1, \\ 2|u_0|/\alpha\xi, & \alpha \gg 1. \end{cases}$$

When $\varepsilon \to -\infty$ from Eq.(20) we find $\varepsilon_l = -|u_0|$.

The root of equations (19) or (20) in zone $\varepsilon \leq \varepsilon_{1\sigma}$ when $u_0 < 0$ corresponds to resonance level $\varepsilon_{r\sigma}$. Its position we derive from equation

$$1 + \frac{8lu_0\alpha^3\xi^2\gamma}{\sqrt{\pi}\varepsilon_0(\alpha+1)|\alpha^2-1|}\left\{\frac{\sqrt{\pi}}{2a\rho} - \frac{\pi k_{1\sigma}}{2}\exp\left(a^2k_{1\sigma}^2\rho^2\right)\left[1-\Phi(ak_{1\sigma}\rho)\right]\right\} = 0, \qquad (23)$$

where

$$k_{1\sigma} = \sqrt{2M(\varepsilon_{1\sigma}-\varepsilon)}.$$

From this equation we obtain that if $ak_{1\sigma}\rho \ll 1$, the resonance level exists only if $|u_0| > u_c$, where

$$u_c = \varepsilon_0 \frac{(\alpha+1)^3\rho}{4\alpha^{7/2}\xi^2}. \qquad (24)$$

Distance $\Delta_1$ between level $\varepsilon_{1\sigma}$ and resonance level in this case is equal to

$$\Delta_1 = \varepsilon_{1\sigma} - \varepsilon_{r\sigma} = \frac{\varepsilon_0}{\pi\rho^2}\left(1 - \frac{u_c}{|u_0|}\right)^2. \qquad (25)$$

Width of this level if $\Delta_1 \gg \omega$ is equal to

$$\Gamma = \varepsilon_0 \frac{(\alpha+1)^2}{\alpha^3\xi^2}\sqrt{\frac{\Delta_1}{\omega}}\exp\left(-\frac{\omega}{\varepsilon_0}\rho^2\right).$$

In particular, when $\alpha \gg 1$ this means that

$$\Gamma = \frac{\varepsilon_0}{\alpha\xi^2}\sqrt{\frac{\Delta_1}{\omega}}\exp\left(-\frac{\omega}{\varepsilon_0}\alpha\xi^2\right). \qquad (26)$$

If $ak_{1\sigma}\rho \gg 1$, the equation (23) results in

$$\Delta_1 = \frac{2|u_0|\alpha^{7/2}\xi^2}{(\alpha+1)^3\rho} = \begin{cases} 2|u_0|\alpha^{7/2}\xi^2, & \alpha \ll 1, \\ \dfrac{2|u_0|}{\alpha\xi}, & \alpha \gg 1. \end{cases} \qquad (27)$$

In an extreme case of $\delta$-potential $a \to 0$, $|u_0| \to \infty$ from (27) this means that

$$\Delta_1 = \frac{\omega|\upsilon_0|}{2\pi\omega_c l^2}.$$

If parabolic potential is absent ($\omega_0 = 0$), then we get

$$\Delta_1 = \frac{|\upsilon_0|}{2\pi l^2}$$

– distance between Landau level and magnetoimpurity level split off from it in a two-dimensional electron gas. Here $l = \left(c/eH\right)^{1/2}$ – magnetic length, $\left(2\pi l^2\right)^{-1}$ – degeneracy order of Landau level. Width of level (27) when $\Delta_1 \ll \omega$ is equal to

$$\Gamma = \frac{\sqrt{\pi}(\alpha+1)^{3/2}\rho\Delta_1^2}{\sqrt{\varepsilon_0\omega}}\exp\left(-\frac{\omega}{\varepsilon_0}\rho^2\right) = \begin{cases} \sqrt{\dfrac{\pi}{\varepsilon_0\omega}}\Delta_1^2\exp\left(-\dfrac{\omega}{\varepsilon_0}\right), & \alpha \ll 1, \\ \sqrt{\dfrac{\pi}{\varepsilon_0\omega}}\alpha^{5/2}\xi^2\Delta_1^2\exp\left(-\dfrac{\omega}{\varepsilon_0}\alpha\xi^2\right), & \alpha \gg 1. \end{cases}$$

Let us list numeric values of functions derived here $\Delta_0$ (22), $u_c$ (24), $\Delta_1$ (25) and $\Gamma$ (26) for parameters $m = 10^{-28}$ g, $|u_0| = 0{,}2*10^{-12}$ erg, $a = 10^{-7}$ cm, $H = 10^4$ Oe, $\omega_0 = \omega_c$, which are typical for the structures studied above. We have $\Delta_0 = 0{,}2*10^{-14}$ erg, $u_c = 0{,}1*10^{-12}$ erg, $\Delta_1 = 0{,}6*10^{-16}$ erg, $\Gamma = 0{,}1*10^{-16}$ erg, $\Gamma/\Delta_1 = 0{,}2$. Thus the local and the resonance levels can be detected for instance in tests aimed at measuring the absorption of electromagnetic radiation in quantum wires at low temperatures in quantizing magnetic field.

### 5. Summary and conclusions

The properties of isolated impurity atoms in massive and low-dimensional conductors are usually studied by the method of local perturbations or by the zero-range radius potential method [1-4,23,26]. Together with these methods in the theory of solid state physics the Keldysh method [18,19] is widely used in combination with functional methods of quantum field theory [16,17]. In particular, this method was used recently to study properties of normal and superconducting metals with randomly distributed impurity atoms [17,20-22]. In papers [20-22] the impurity scattering of electrons is taken into account within Born approximation. Meanwhile in two- and one-dimensional structures the impurity donors of arbitrary small strength are capable to form local and resonant states of electrons. They will be formed also in massive conductors at presence of a magnetic field [6,7,10,11]. That is why there exist a necessity to go beyond the Born approximation and to take into account these states within the framework of Keldysh formalism. In this paper we have shown that the characteristics of impurity states of electrons in nanostructures in a magnetic field can be evaluated by functional methods. The theory is applied to electrons in quantum wires. The model of quantum wires is the two-dimensional electron gas with parabolic confinement potential. The impurity atom field is modelled by a Gaussian separable potential. The positions and the widths of local and resonant levels of energy of electrons caused by combined effect of donor impurity atoms and magnetic field are obtained. Functional approach combined with Keldysh formalism allows us to study the influence of magnetoimpurity states of electrons on kinetic properties of non-equilibrium conductors taking into account the interaction of electrons among themselves and with other quasi-particles, as well as the influence on weak localization of electrons in magnetic field.

## Acknowledgments

This work is partially supported by INTAS program (grant INTAS-01-0791). We would like to thank T. Rashba for help with the preparation of the manuscript.